\documentclass[a4paper,11pt]{article}
\usepackage{pos}

\usepackage{subfigure}

\newcommand{\LL}{\mathcal{L}}
\newcommand{\II}{\mathcal{I}}
\newcommand{\dd}{\mathrm{d}}
\newcommand{\ord}{\mathcal{O}}
\newcommand{\stat}{{\mathrm{stat}}}
\newcommand{\sys}{{\mathrm{sys}}}
\newcommand{\hlbl}{{\mathrm{hlbl}}}
\newcommand{\tpt}{(2\!+\! 2)}
\newcommand{\fig}{figure\ }
\newcommand{\Fig}{Figure\ }
\newcommand{\tab}{table\ }

\title{The hadronic light-by-light contribution to the muon $g{-}2$ using staggered fermions at the physical point}
\ShortTitle{HLbL contribution to the muon $g{-}2$}

\author*[a]{Christian Zimmermann}
\author[a]{Antoine G\'{e}rardin}

\affiliation[a]{Aix Marseille Univ, Université de Toulon, CNRS, CPT, Marseille, France}

\emailAdd{christian.zimmermann@univ-amu.fr}
\emailAdd{antoine.gerardin@univ-amu.fr}

\abstract{Hadronic contributions dominate the uncertainty of the Standard Model prediction for the anomalous magnetic moment of the muon. In this work, we present results on the hadronic light-by-light contribution obtained from the evaluation of the hadronic four-point function of electromagnetic currents using the position-space formalism developed by the Mainz group. The simulations are performed with staggered fermions directly at the physical point. Several physical volumes are used to estimate finite volume effects. This direct lattice study is supplemented by considering the contribution of the light pseudoscalar pole in both finite and infinite volumes, where we reuse the pseudoscalar transition form factors that have been evaluated in previous simulations on the same ensembles.}

\FullConference{The 41st International Symposium on Lattice Field Theory (LATTICE2024)\\
 28 July - 3 August 2024\\
Liverpool, UK\\}

\begin{document}
\maketitle

\section{Introduction}

Hadronic contributions remain the major source of uncertainties in the Standard Model (SM) determination of the anomalous magnetic moment of the muon, which is quantified by $a_\mu = (g_\mu - 2)/2$. Since uncertainties on the experimental side are constantly reduced by ongoing experiments \cite{Muong-2:2023cdq}, theoretical calculations need to be improved as well, in order to make suitable comparisons and discover potential hints of new physics. A summary of the current state of research is given by \cite{Aoyama:2020ynm}. There are two main contributions in the hadronic sector, the hadronic vacuum polarization (HVP) and the hadronic light-by-light (HLbL) scattering. Both can be studied on the lattice, see \cite{Chao:2021tvp,Chao:2022xzg,Blum:2019ugy,Blum:2023vlm} for the HLbL contribution.

In these proceedings, we summarize the current status of our research regarding the direct calculation of the HLbL contribution by evaluating the corresponding four-point correlation on the lattice. More details can be found in \cite{Fodor:2024jyn}.

\section{Lattice calculation with staggered fermions}

\begin{figure}
\begin{center}
\includegraphics[scale=0.2]{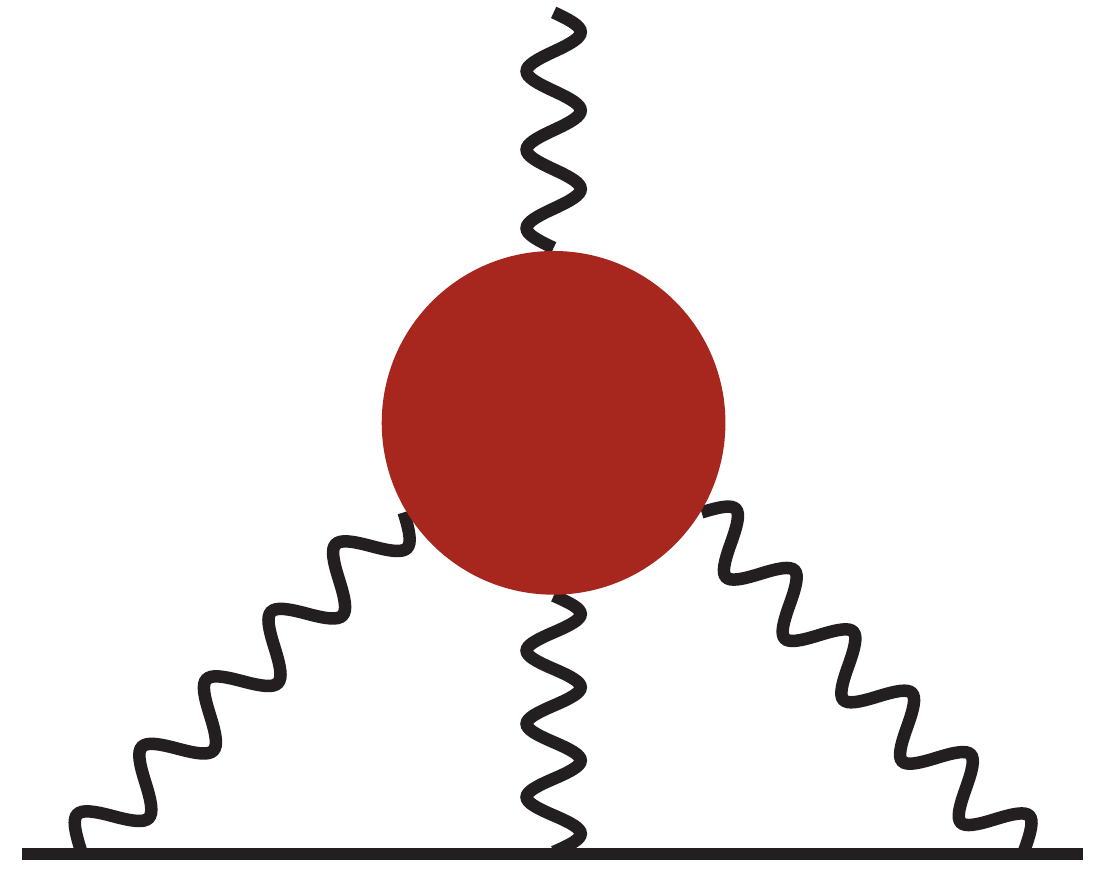}
\end{center}
\vspace*{-0.6cm}
\caption{Feynman diagram representing the hadronic light-by-light (HLbL) contribution. The red blob corresponds to the hadronic part.\label{fig:hlbl}}
\end{figure}

Hadronic light-by-light (HLbL) scattering contributes to the anomalous magnetic moment of the muon at $\ord(\alpha^3)$ of the fine structure constant $\alpha$. The corresponding Feynman diagram is shown in \fig\ref{fig:hlbl}. Its value is given by the master formula \cite{Chao:2020kwq,Chao:2021tvp}
\vspace*{-0.4cm}

\begin{align}
a_\mu^{\mathrm{hlbl}} 
&= 
-\frac{m_\mu e^6}{3} \int \dd^4 y \int \dd^4 x\ \LL^{\mathrm{sym}}_{[\rho,\sigma];\mu\nu\lambda}(x,y) \int \dd^4 z\ z_\rho\ \Pi_{\mu\nu\sigma\lambda}(x,y,z) 
\nonumber\\
&=
2\pi^2 \int_{|y|} |y|^{3} f(y) = \int_{|y|} \II(|y|) \,,
\label{eq:hlbl}
\end{align}
where $\LL^{\mathrm{sym}}$ is a symmetrized version of the QED kernel $\LL^{(\Lambda)}$ with $\Lambda = 0.4$ \cite{Asmussen:2022oql}, which represents the photon lines and the muon line in \fig\ref{fig:hlbl}. The hadronic part depicted by the red blob is given by the four-current correlation function, which reads
\vspace*{-0.4cm}

\begin{align}
\Pi_{\mu\nu\sigma\lambda}(x,y,z) &= \langle j_\mu(x) j_\nu(y) j_\sigma(z) j_\lambda(0) \rangle \,.
\label{eq:4pt-function}
\end{align}
The electromagnetic current $j_\mu(x)$ takes into account contributions by the light, strange and charm quarks. After executing the integrals over $x$ and $z$, the remaining integrand $f(y)$ is Lorentz invariant. Therefore, it is sufficient to sample the integrand for only a few values of $|y|$. The integral over the 3-sphere is done by taking into account the factor $2\pi^2 |y|^3$. We denote the weighted integrand by $\mathcal{I}(|y|)$.

\begin{figure}
\begin{center}
\includegraphics[scale=1]{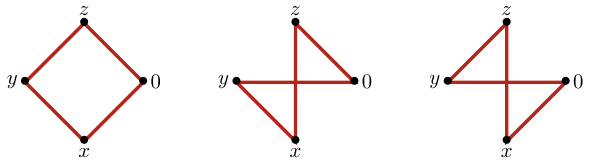}
\end{center}
\vspace*{-0.6cm}
\caption{The three connected Wick contractions contributing to the four-current correlator \eqref{eq:4pt-function}. We denote the corresponding contractions $\Pi^{\mathrm{conn},(1)}$, $\Pi^{\mathrm{conn},(2)}$, $\Pi^{\mathrm{conn},(3)}$ from left to right.\label{fig:conn}}
\end{figure}

\begin{figure}
\begin{center}
\includegraphics[scale=1]{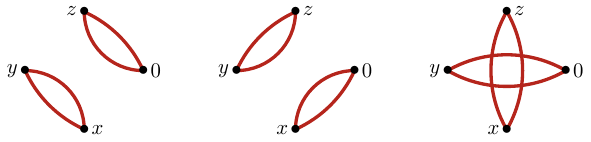}
\end{center}
\vspace*{-0.6cm}
\caption{The leading disconnected (2+2) diagrams contributing to the four-current correlator \eqref{eq:4pt-function}. We denote the corresponding contractions $\Pi^{(2+2),(1)}$, $\Pi^{(2+2),(2)}$, $\Pi^{(2+2),(3)}$ from left to right.\label{fig:2+2}}
\end{figure}

In the context of a lattice simulation, the four-current correlation function decomposes in 24 Wick contractions, which can be grouped in five types. These are classified by the number of currents connected by quark lines in each quark-disconnected part. In these proceedings we only consider connected diagrams and leading disconnected (2+2) diagrams. For the first kind, there are six different diagrams appearing pairwise with inverted fermion flow. They are depicted in \fig\ref{fig:conn}. For the (2+2) contribution, there are three different diagrams shown in \fig\ref{fig:2+2}

In order to reduce the computational cost, we use translational invariance to reduce two of the connected diagrams to the diagram $\Pi^{\mathrm{conn},(1)}$, where no sequential inversions are needed:
\vspace*{-0.4cm}

\begin{align}
a_\mu^{\mathrm{conn}} = \frac{m_\mu e^6}{3} 2\pi^2 \sum_{|y|} |y|^{3} \sum_{x,z} \LL^{\mathrm{sym}}_{[\rho,\sigma];\mu\nu\lambda}(x,y)\ (x_\rho - 3 z_\rho)\ \Pi^{\mathrm{conn},(1)}_{\mu\nu\sigma\lambda}(x,y,z) \,.
\end{align}
A similar approach is done for the (2+2) contribution. In this case, there are two versions, where all diagrams are reduced to either $\Pi^{(2+2),(1)}$ or $\Pi^{(2+2),(2)}$. In the end, we take the average of both versions in order to improve statistics.
%
%

The present calculations are performed using staggered fermions. The sums over $x$ and $z$ in \eqref{eq:hlbl} are performed explicitly in the lattice simulation, so that we are left with a Lorentz invariant quantity that depends only on $|y|$. However, since this is a position space quantity, the presence of staggered fermions leads to contaminations from 16 taste partners. In order to project onto the taste singlet, we apply a smearing function w.r.t.\ $y$ to the quantity $f(y)$ (see \eqref{eq:hlbl}). For more details, see \cite{Zimmermann:2023tal,Fodor:2024jyn}.

\section{Simulation and Results}

The present calculations are carried out using ensembles generated by the BMW collaboration \cite{Borsanyi:2020mff,Boccaletti:2024guq}. The ensembles employ $N_f=2+1+1$ dynamical staggered fermions with four steps of stout smearing. The pseudoscalar masses correspond to the physical ones. The simulations are performed for three different volumes ($L=3,4,6~\mathrm{fm}$) and for lattice spacings $0.06~\mathrm{fm}\le a \le 0.13~\mathrm{fm}$. For the light quark contribution we focus on the two larger lattice volumes.

\subsection{Light quark contribution}

\begin{figure}
\begin{center}
\subfigure[$a=0.1097~\mathrm{fm}$, $L=3~\mathrm{fm}$]{
\includegraphics[scale=0.35, clip, trim=0 0.55cm 0 0.55cm]{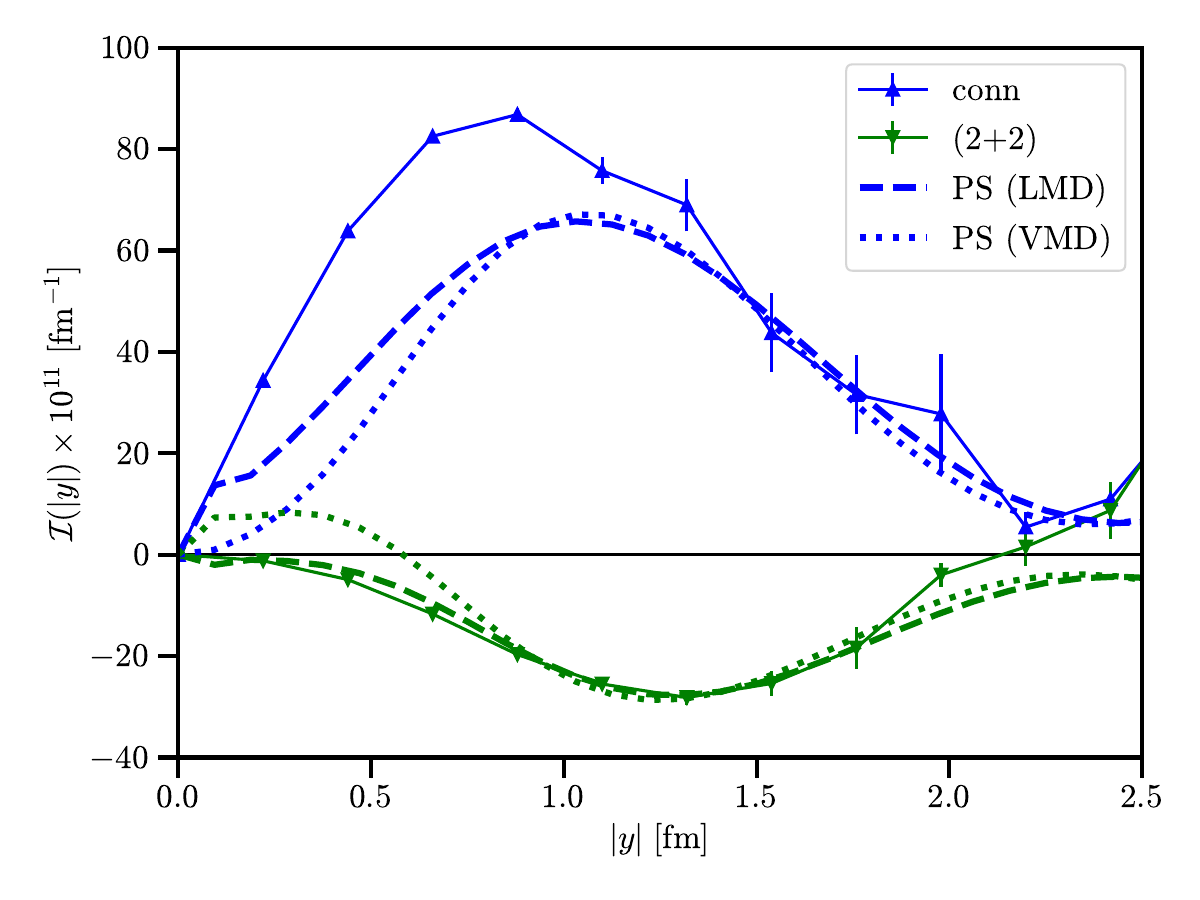}
}
\subfigure[$a=0.1097~\mathrm{fm}$, $L=6~\mathrm{fm}$]{
\includegraphics[scale=0.35, clip, trim=0 0.55cm 0 0.55cm]{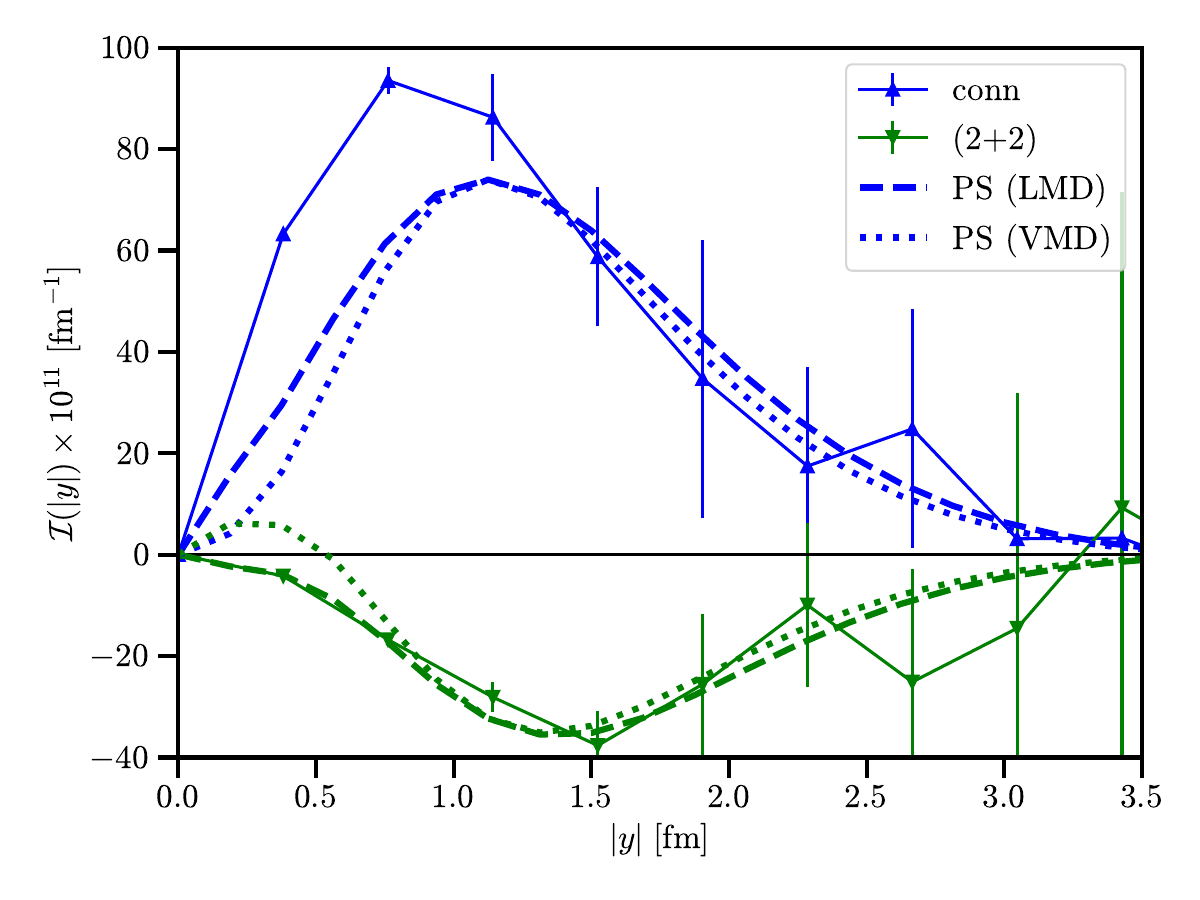}
}
\end{center}
\vspace*{-0.6cm}
\caption{Integrand for the connected (blue) and 2+2 (green) contribution for $a=0.1097~\mathrm{fm}$ and $L=3~\mathrm{fm}$ (a) or $L=6~\mathrm{fm}$ (b), respectively.\label{fig:light-intnd}}
\end{figure}

\begin{figure}
\begin{center}
\subfigure[Contribution of the tail.\label{fig:tail-impr}]{
\includegraphics[scale=0.35, clip, trim=0 0.55cm 0 0.55cm]{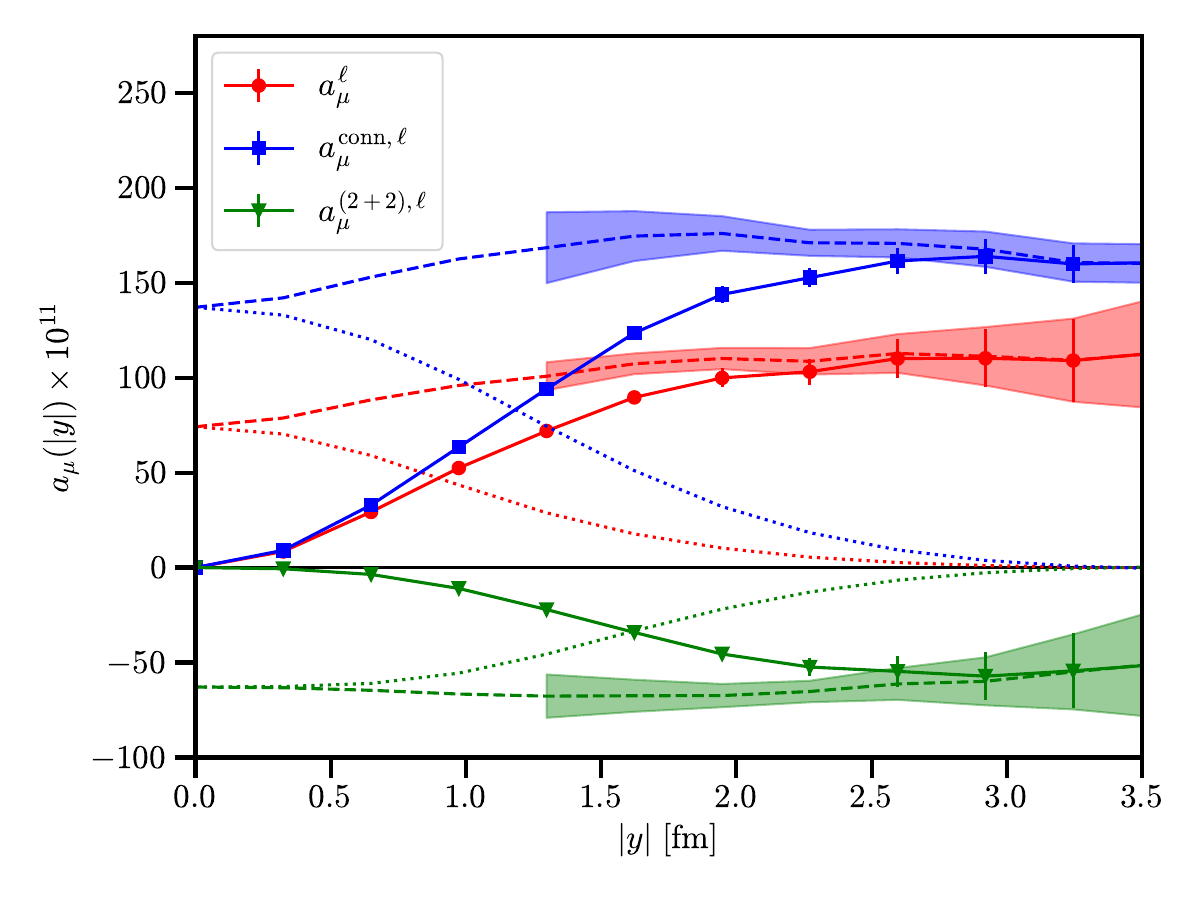}
}
\subfigure[Continuum extrapolation.\label{fig:light-cont}]{
\includegraphics[scale=0.35, clip, trim=0 0.55cm 0 0.55cm]{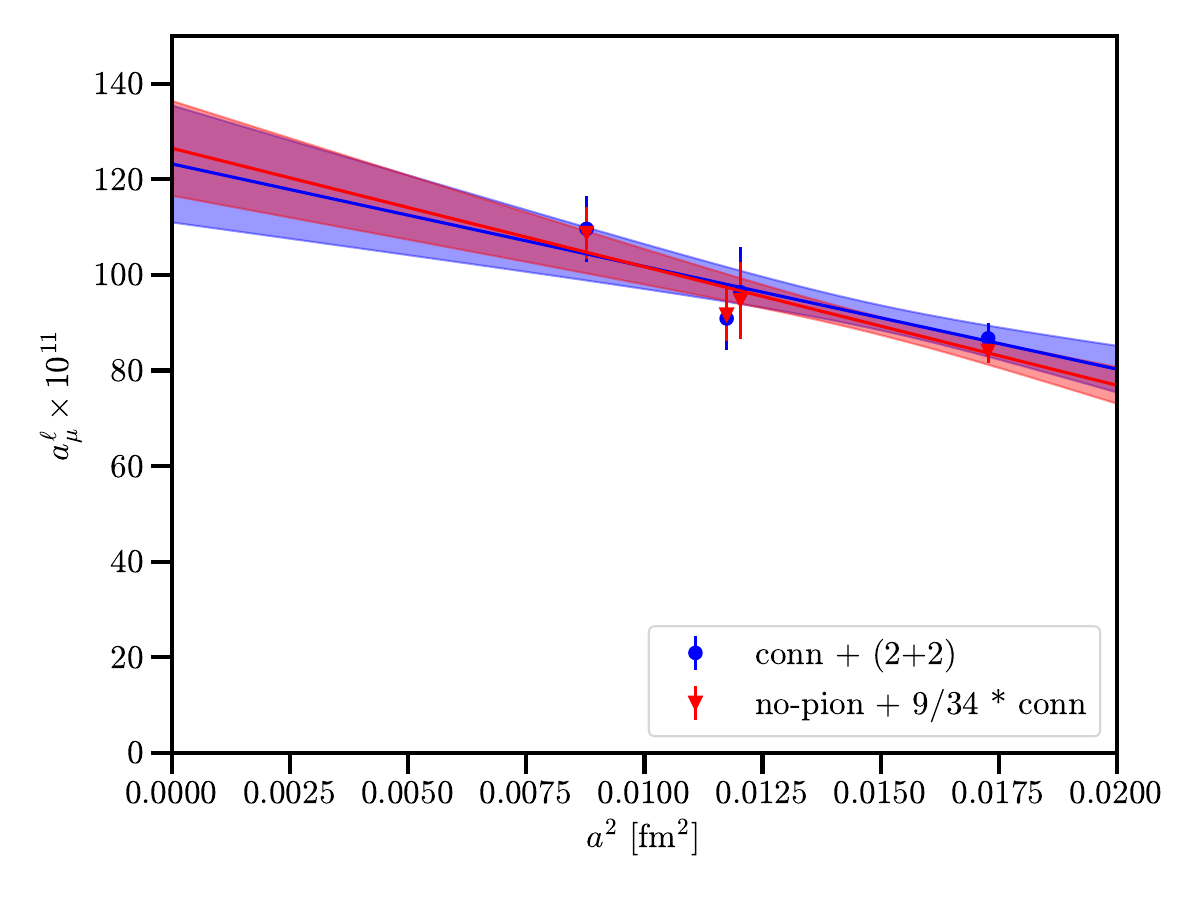}
}
\end{center}
\vspace*{-0.6cm}
\caption{Left: The dashed line shows the result for \eqref{eq:tailcut} (red) depending on $|y|_\mathrm{cut}$. Analogous results are shown for the connected (blue) and (2+2) (green) contributions individually. The solid line shows the corresponding contributions from the lattice data, while the dotted line is the pseudoscalar-pole contribution. Right: Continuum extrapolation for the total light contribution.\label{fig:light-plots}}
\end{figure}

\Fig\ref{fig:light-intnd} shows the integrand of the connected (blue) and (2+2) contributions (green) for $L=3~\mathrm{fm}$ (left) and $L=6~\mathrm{fm}$ (right), both for lattice spacing $a\approx 0.11~\mathrm{fm}$. The long distance region is dominated by contributions from the pseudoscalar poles $\pi^0$, $\eta$ and $\eta^\prime$. Hence we can use input from lattice calculations of the corresponding transition form factors (TFFs) to improve our lattice results. The corresponding calculation has been performed on the same ensembles as in the present study \cite{Gerardin:2023naa}. To this end, we write the contribution to the four-current correlation function \eqref{eq:4pt-function} by the pole $P$ and denote the corresponding integrand by $\II^P(|y|)$. The functional form of the TFFs is approximated by light meson dominance (LMD) or vector meson dominance (VMD) models, respectively. For details on that, see \cite{Fodor:2024jyn}. The integrand calculated from the pseudoscalar pole data is also plotted in \fig\ref{fig:light-intnd} (dashed lines for LMD, dotted for VMD). It matches perfectly for large distances with the four-point data for both volumes.
%
%
%
The pseudoscalar pole contribution can be employed in a two-fold way. First we can estimate finite volume corrections of the light quark contribution for a given volume:
\vspace*{-0.4cm}

\begin{align}
v^{P} = a_\mu^{P}(\infty) - a_\mu^P(V) \,.
\end{align}
Second, we replace the original four-point function integrand for large distances $|y| > |y|_\mathrm{cut}$ in a manner that the overall pseudoscalar pole contribution is small compared to the statistical error:
\vspace*{-0.4cm}

\begin{align}
a_\mu^{\ell} =\! \int_0^{|y|_{\mathrm{cut}}}\!\! \dd |y| \left[ \II^{\mathrm{conn}}(|y|)\! +\! \II^{(2+2)}(|y|) \right] +\! \int_{|y|_{\mathrm{cut}}}^\infty\!\! \dd |y| \left[ \II^{\pi}(|y|)\! +\! \II^{\eta}(|y|)\! +\! \II^{\eta^\prime}(|y|) \right] +\! v^{\pi}\! + v^{\eta}\! + v^{\eta^\prime} .
\label{eq:tailcut}
\end{align}
Similar approaches are derived for the connected and (2+2) contributions individually. \Fig\ref{fig:tail-impr} shows the result of \eqref{eq:tailcut} depending on $|y|_\mathrm{cut}$ (dashed lines). The contribution of the lattice data (solid lines) and the pseudoscalar integrands (dotted lines) are also shown. An alternative to \eqref{eq:tailcut} is given by calculating 
\vspace*{-0.4cm}

\begin{align}
a_\mu = a_\mu^{\mathrm{no-}\pi} + \frac{9}{34} a_\mu^{\mathrm{conn},\ell} + v^{\pi} + v^{\eta} + v^{\eta^\prime} \,,
\label{eq:nopi}
\end{align}
where $a_\mu^{\mathrm{no-}\pi}$ is obtained from the integrand $(25/34) \II^{\mathrm{conn},\ell}(y) + \II^{(2+2)}(y)$ \cite{Gerardin:2017ryf}. This has already been used in \cite{Blum:2023vlm}. This has the advantage that the long-range pion-pole contribution is removed and, therefore, the integral converges faster, so that the statistical error is smaller.

For the continuum extrapolation we take into account four ensembles for $L=4,6~\mathrm{fm}$ at three different lattice spacings. This allows us to perform a two parameter fit using the following ansatz:
\vspace*{-0.4cm}

\begin{align}
a_\mu^{\ell}(a) = a_\mu^{\mathrm{cont},\ell} + \beta_2 (\Lambda a)^2 \,.
\end{align}
Our result reads $a_\mu^{\ell} = 122.6(11.6)\times 10^{-11}$ for the direct calculation and $a_\mu^{\ell} = 126.2(9.2)\times 10^{-11}$ if we use the alternative approach \eqref{eq:nopi}. The two extrapolations are plotted in \fig\ref{fig:light-cont}

\subsection{Strange quark contribution}

\begin{figure}
\begin{center}
\subfigure[\label{fig:strange-intnd}]{
\includegraphics[scale=0.25]{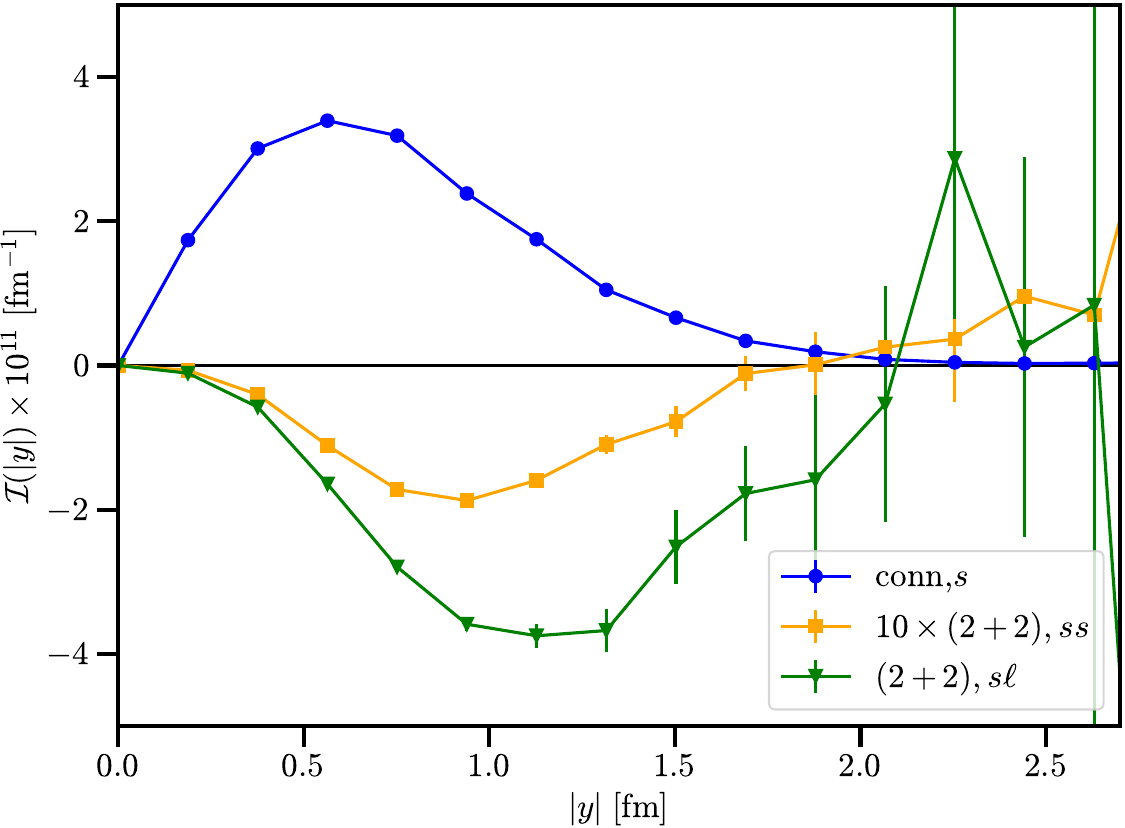}
}
\subfigure[\label{fig:strange-conn-cont}]{
\includegraphics[scale=0.25]{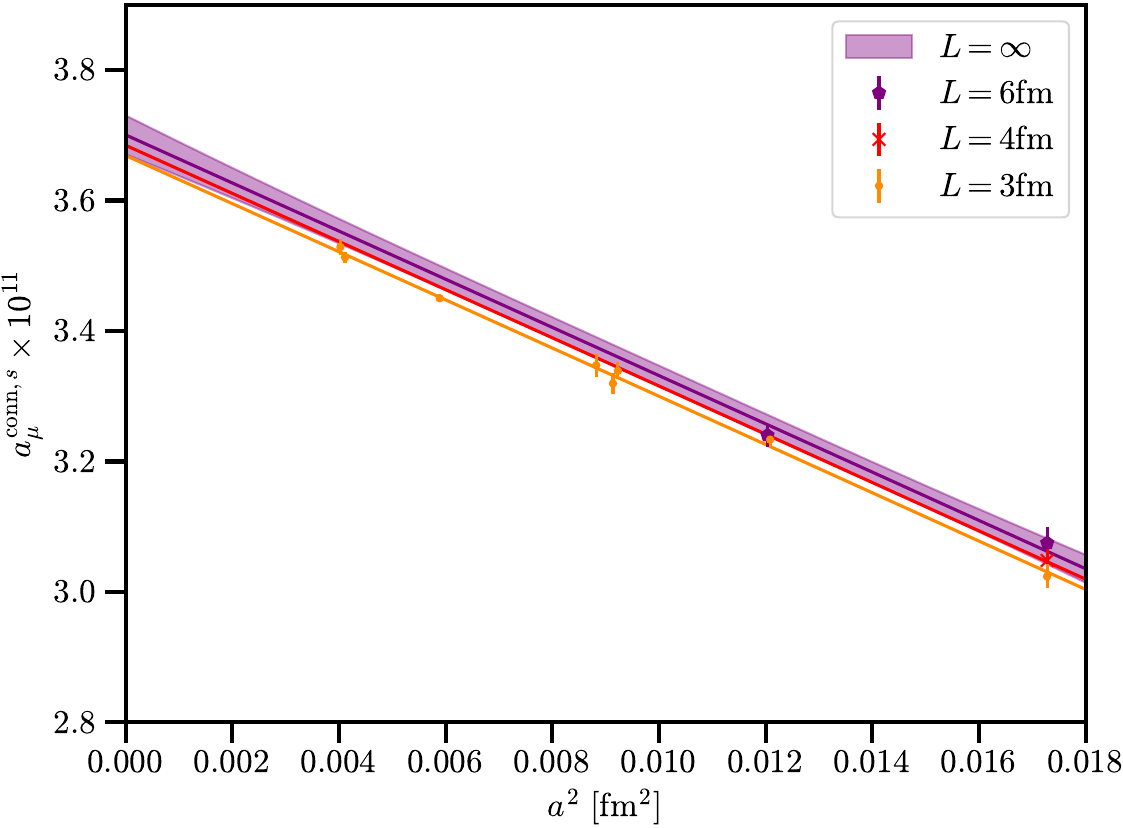}
}
\subfigure[\label{fig:strange-2p2-cont}]{
\includegraphics[scale=0.25]{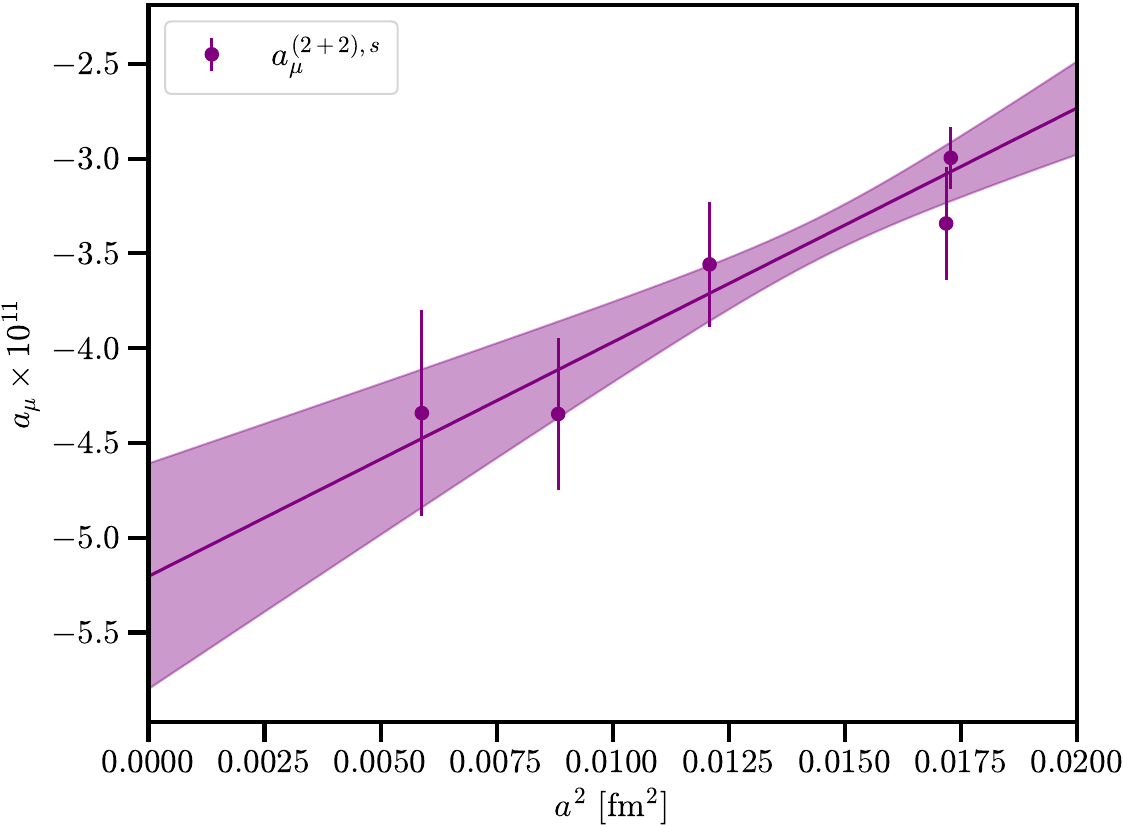}
}
\end{center}
\vspace*{-0.7cm}
\caption{(a) The integrand for the strange quark contributions. For visibility, the strange-strange $\tpt$-contribution is multiplied by a factor of 10. (b,c) Continuum extrapolation for the strange connected contribution (b) and the strange $\tpt$-contribution (c).\label{fig:strange-cont}}
\end{figure}

The integrand for the strange quark contributions is shown in \fig\ref{fig:strange-intnd} for $L=3~\mathrm{fm}$ and $a=0.094~\mathrm{fm}$. The strange contribution is dominated by the connected diagrams and by the light-strange (2+2) contributions. The latter has negative sign and, therefore, more or less cancels the signal of the connected diagram. The strange connected contribution has very small statistical errors. Hence lattice artifacts and finite volume effects become non-negligible.

As ansatz for the continuum extrapolation we generally consider $a^2$ and $a^4$ terms, as well as a scale dependence introduced by powers of $\alpha_s(a^{-1})$. Moreover, we have to take into account the slight miss-tuning of the strange quark mass. Here we use the ratio $\delta M_{ss}/M_{ss}^{\mathrm{phys}}$ as proxy, where $M_{ss}^{\mathrm{phys}} = 689.89(49)~\mathrm{MeV}$ and $\delta M_{ss}$ is the difference obtained for the given ensemble. The finite volume effects are taken into account by the exponential term $e^{-L m_\pi}$, where $m_\pi$ is the physical pion mass:
\vspace*{-0.4cm}

\begin{align}
a_\mu^{s}(a) = a_\mu^{\mathrm{cont},s} + \beta_2 (\Lambda a)^2 + \beta_4 (\Lambda a)^4 + \delta_2 (\Lambda a)^2 \alpha_s^n (a^{-1}) + \gamma \frac{\delta M_{ss}}{M_{ss}^{\mathrm{phys}}} + \lambda e^{-L m_\pi} \,.
\end{align}
For the strange connected contribution, we take into account 11 ensembles for 3 different volumes. We perform different fits where $\beta_4$ or $\delta_2$ or both are set to zero. These fits are used to estimate the systematic error. For the same reason we vary the fit range. An exemplary fit is plotted in \fig\ref{fig:strange-conn-cont}. As our final result we state $a_\mu^{\mathrm{conn},s} = 3.694(25)_\stat(8)_\sys\times 10^{-11}$.

In the case of the (2+2) contribution, where statistical noise is larger, we can neglect the miss-tuning and finite volume effects, and perform a two-parameter fit. The result is plotted in \fig\ref{fig:strange-2p2-cont}. Our continuum value reads $a_\mu^{(2+2),s} = -5.4(0.8)_\stat(0.2)_\sys\times 10^{-11}$.

\section{Conclusion}

\begin{table}
\begin{center}
\begin{tabular}{cc}
\hline
\hline
Contribution & $a_\mu \times 10^{11}$ \\
\hline
Light total & $122.6(11.6)_\stat$ \\
Strange total & $-1.7(0.8)_\stat(0.3)_\sys$ \\
Charm total & $3.73(5)_\stat(26)_\sys$ \\
Sub-leading disc. & $0.83(25)_\stat$ \\
\hline
Total & $125.5(11.6)_\stat(0.4)_\sys$ \\
\hline
\hline
\end{tabular}
\end{center}
\vspace*{-0.6cm}
\caption{Compilation of our preliminary results for all individual contributions to $a_\mu^\hlbl$. \label{tab:results}}
\end{table}

We considered all Wick contractions contributing to the hadronic light-by-light scattering contribution to the anomalous magnetic moment using staggered fermions and at physical masses. In these proceedings, we focus only on connected and (2+2) diagrams for the light and strange quark contributions. The light quark contribution has been supplemented by estimates of the tail obtained from pseudoscalar transition form factor results. This is done in a way so that systematics remain small compared to the statistical error. The light quark contribution has been found to be the largest contribution. The result for the strange contribution is even small compared to the target precision of $10\%$ due to large cancellations between connected in disconnected diagrams. A compilation of results for the individual contribution is shown in \tab\ref{tab:results} including our preliminary results for the charm and sub-leading contributions, which have not been discussed in detail in these proceedings. Our result for the hadronic light-by-light contribution to the anomalous magnetic moment of the muon reads:
\vspace*{-0.4cm}

\begin{align}
a_\mu^\hlbl = 125.5(11.6)_\stat(0.4)_\sys \times 10^{-11}\,.
\end{align}

This result is compatible with recent results of other lattice collaborations \cite{Chao:2021tvp,Chao:2022xzg,Blum:2019ugy,Blum:2023vlm}. At the current stage, the overall uncertainty is around $10\%$.

We plan to repeat our analysis of the light quark contribution for another ensemble with lattice spacing $a\approx 0.077~\mathrm{fm}$ in order to improve the quality of our continuum extrapolation and, therefore, to further reduce our overall uncertainty.

\section*{Acknowledgments}

We thank all the members of the Budapest-Marseille-Wuppertal collaboration for helpful discussions and the access to the gauge ensembles used in this work. The configurations were generated from computer times provided by the Gauss Centre for Supercomputing on the machines JUWELS, SUPERMUC and HAWK. This publication received funding from the Excellence Initiative of Aix-Marseille University - A*Midex, a French ``Investissements d'Avenir" programme, AMX-18-ACE-005 and from the French National Research Agency under the contract ANR-20-CE31-0016.
The computations were performed on Irene at TGCC. We thank GENCI (grants A0080511504, A0100511504 and A0120511504) for awarding us computer time on these machines.
Centre de Calcul Intensif d'Aix-Marseille (CCIAM) is acknowledged for granting access to its high performance computing resources.


\bibliographystyle{JHEP}
\setlength{\bibsep}{1pt}
\bibliography{biblio}

\end{document}